\documentclass[a4paper,english,10pt,superscriptaddress,reprint,amsmath,amssymb,aps,twocolumn,floatfix,notitlepage,showkeys,prd]{revtex4-2}
\usepackage{graphicx}
\usepackage{bm}
\usepackage{dcolumn}
\usepackage[colorlinks=true,citecolor=darkred,urlcolor=darkred, pdfborder={0 0 0}]{hyperref}
\usepackage{setspace}
\usepackage{caption}
\usepackage{subcaption} 
\usepackage{slashed}
\usepackage[normalem]{ulem}
\usepackage{float}
\usepackage[usenames,dvipsnames]{xcolor}  
\usepackage{microtype}
\usepackage{lipsum}
\usepackage{amsmath}
\usepackage{amssymb}
\usepackage{mathrsfs}
\usepackage{placeins}
\usepackage{lettrine}   
\input Eileen.fd        
\newcommand*\initfamily{\usefont{U}{Eileen}{xl}{n}}

\definecolor{darkred}{rgb}{0.6,0,0}
\definecolor{linkcolor}{rgb}{0,0,0.5}
\usepackage[T1]{fontenc} 
\usepackage[compat=1.1.0]{tikz-feynhand}
\usepackage{tikz-feynman}
\tikzfeynmanset{compat=1.1.0}
\usepackage{feynmp}
\usepackage{tikzsymbols}
\usepackage{array}
\usepackage{pifont} 

\definecolor{linkcolor}{rgb}{0,0,0.5}
\usepackage{orcidlink}
\begin{document}
\preprint{2507.10299}
%
\title{\boldmath \color{BrickRed}Scotogenic Model with Non-invertible Symmetry}
%
\author{Takaaki Nomura \orcidlink{0000-0002-0864-8333}}
\email{nomura@scu.edu.cn}
\affiliation{College of Physics, \href{https://ror.org/011ashp19}{Sichuan University}, Chengdu 610065, China}
\author{Oleg Popov \orcidlink{0000-0002-0249-8493}}
\email{opopo001@ucr.edu (corresponding author)}
\affiliation{Faculty of Physics and Mathematics, \href{https://ror.org/02q963474}{Shenzhen MSU-BIT University},
Shenzhen 518172, China}
\date{\today}
%
\begin{abstract}
In the present work, the scotogenic model is constructed applying non-invertible $Z_M$ symmetries. The stability of dark matter and the scotogenic structure of the neutrino mass matrix is achieved via a novel application of non-invertible symmetry. The Scotogenic model with non-invertible symmetry is given with minimalistic content giving a one-zero structure of the neutrino mass matrix, and numerical analysis of the lepton mixing angles and physics are presented. Other relevant constraints are also studied.
\end{abstract}
\keywords{neutrino, scotogenic, no-group, non-invertible, lepton mixing}
\maketitle 
%
\section{Introduction}
\label{sec:intro}
\lettrine[lines=4,findent=-1.2cm]{\normalfont\initfamily \fontsize{17mm}{10mm}\selectfont N \normalfont\initfamily}{}eutrino masses and mixing angles have been a puzzle to physicists for the recent 30 years, since the neutrino oscillations observation in the 90's. The origin of neutrino masses, which requires Beyond the Standard Model (BSM) physics, has been a cherished goal for many. There have been many proposals on how neutrino masses are produced. Some are minimalistic, some are more complicated than others, and others connect neutrino mass origin with yet another to be confirmed BSM physics. Among such proposals stands the scotogenic mechanism that connects the origin of naturally small neutrino masses with the existence of dark matter\cite{Ma:2006km}.

In regard to the lepton mixing angles that are measured via the various neutrino oscillations, there is an extensive study list that covers everything starting from discrete symmetries~\cite{Babu:2002dz,Ma:2001dn,Altarelli:2010gt,Hernandez:2012ra,King:2013eh,King:2014nza,Kobayashi:2022moq}, which has been an orthodox approach to the explanation of the sizes of the lepton mixing angles. The lion's share of the study lies in this direction; then comes a modular symmetries~\cite{Feruglio:2017spp,Kobayashi:2023zzc,Ding:2023htn,Kobayashi:2018vbk,Nomura:2019jxj,Nomura:2019lnr,Dasgupta:2021ggp,Kobayashi:2023qzt}, that gained some attention in the last decade or so and have shown success to a certain point; in the later years a non-SUSY version of the modular symmetries, \emph{a.k.a.} non-holomorphic modular approach~\cite{Qu:2024rns,Nomura:2024ctl,Li:2024svh,Nomura:2025ovm,Nomura:2024vzw}, takes place which is more minimalistic compared to modular version and requires no super partners.

Finally, a more recent novel detour of non-invertible (\emph{a.k.a.} no-group) symmetries, see~\cite{Gomes:2023ahz,Schafer-Nameki:2023jdn,Bhardwaj:2023kri,Shao:2023gho} for review and~\cite{Kobayashi:2024yqq,Kobayashi:2024cvp,Delgado:2024pcv,Funakoshi:2024uvy,Kobayashi:2025znw,Liang:2025dkm,Kobayashi:2025ldi,Kobayashi:2025cwx,Kobayashi:2025lar,Nomura:2025sod,Dong:2025jra,Suzuki:2025oov,Cordova:2022fhg} for applications. This approach, unlike others, has a different way of building selection rules, compared to orthodox approach which takes invariants of group actions into selection rules, here the conjugacy class invariants are used to construct Lagrangian invariants. Remarkably, this kind of symmetries is broken by radiative correction even if it is exact at tree level see~\cite{Heckman:2024obe,Kaidi:2024wio,Suzuki:2025oov,Kobayashi:2025cwx}, where systematic studies associated with radiative neutrino mass generation are discussed in ref.~\cite{Kobayashi:2025cwx}. This scheme offers new possibilities and predictions for BSM physics. Then we discuss application of such a symmetry to a scotogenic model. The $Z_2$ symmetry in the canonical scotogenic model is \textit{ad hoc}, whereas in non-invertible $Z_M$ scenario, the $Z_M$ that serves the purpose of stabilizing dark matter (scotogenic symmetry) has its roots in string theory~\cite{Kobayashi:2024yqq}.

In addition to the discrete symmetry solutions of the lepton flavor mixing, there are also continuous symmetry (local or global) explanations of the flavor mixing in the leptonic sector~\cite{Froggatt:1978nt,Altarelli:2004za,Arcadi:2022ojj,Calibbi:2025rxn,Giarnetti:2025idu,Linster:2018avp,Petcov:1982ya,Petcov:2004rk,Linster:2020fww,Barbieri:1995uv,Barbieri:1997tu,Roberts:2001zy,Darme:2023nsy,Greljo:2023bix}.

The next section describes the evolved scotogenic model with conventional \emph{ad.hoc.} $Z_2$ discrete group symmetry replaced by non-invertible $Z_M$ symmetry. It also explains the details of the non-invertible symmetry mechanics. In section~\ref{sec:num_ana} the details of the numerical calculations and results are given. Finally, section~\ref{sec:discussion} contains a discussion and closing remarks on the present work.

\section{Model setup}
\label{sec:model}

In this section, we discuss the setup of a scotogenic model with a non-invertible symmetry. 
In the scenario the discrete $Z_2$ symmetry in the original scotogenic model is converted into a non-invertible $Z_M$ symmetry.
As in the original scotogenic model, we introduce the second Higgs doublet $\eta$ and SM singlet fermions $N_\alpha$.
The resultant model based on non-invertible $Z_M$ symmetry is minimalistic in a sense that it does not require any new fields beyond the canonical scotogenic model. The only update with respect to canonical scotogenic model is the replacement of scotogenic discrete $Z_2$ symmetry by novel non-invertible $Z_M$ symmetry.
The relevant Lagrangian for the scotogenic model is given by
\begin{equation}
\mathcal{L}_{Y_\ell} = y^e \overline{L} \ell H + y^\eta \overline{L} N \eta + \frac12 M_N \overline{N^c} N + h.c.,
\end{equation}
where $h.c.$ indicates the hermitian conjugate.
In addition, the scalar potential of the model is given by
\begin{align}
V & = \mu_\eta^2  \eta^\dagger \eta - \lambda_0 (H \eta)^2  + \lambda_{H\eta} \left(H^\dagger H\right)\left(\eta^\dagger \eta\right) \nonumber \\
&  + \lambda_{H\eta}^\prime \left|H \eta\right|^2 + \text{other trivial terms}.
    \end{align}
The quadratic term $H \eta$ is forbidden by a non-invertible $Z_M$ symmetry guaranteeing the vanishing vacuum expectation value (VEV) of $\eta$.
The electroweak symmetry is spontaneously broken in the same way as in the SM after the SM Higgs field develops its VEV as $\langle H \rangle =v/\sqrt{2}$.
We write the inert scalar doublet $\eta$ by
\begin{equation}
\eta = \begin{pmatrix} \frac{1}{\sqrt2} (\eta_R + i \eta_I) \\ \eta^- \end{pmatrix},
\end{equation}
where $\eta^-$ has an electric charge $-1$ and the other components are electrically neutral.
After electroweak symmetry breaking, squared masses of inert scalar bosons are
\begin{subequations}
    \label{eq:eta_masses}
    \begin{align}
        m_{\eta_{R[I]}}^2 &= \mu^2_\eta + \frac{v^2}{2}(\lambda_{H \eta} + \lambda'_{H \eta} -[+] 2 \lambda_0), \\
        m_{\eta^\pm}^2 &= \mu^2_\eta + \frac{v^2 \lambda_{H \eta}}{2},
    \end{align}
\end{subequations}
where we write $\Delta m^2 \equiv  m^2_{\eta_I} - m^2_{\eta_R} = 2 \lambda_0  v^2$.

\subsection{Minimal assignment}
We search for a minimal choice of $M$ for $Z_M$ and assignments of classes $[g^{k}]$ for leptons/scalars under the following criteria:
\begin{itemize}
\item The flavors of lepton are distinguished by class $[g^k]$.
\item The neutrino mass is generated only through the scotogenic mechanism at the one loop level.
\item The stability of dark matter is guaranteed, at least at a renormalizable level, by the symmetry. 
\item The SM Higgs field is trivial under the $Z_M$ so that the quark sector is the same as the SM.
\end{itemize}
The definition of classes and selection rules under the non-invertible $Z_M$ symmetries are summarized in the Appendix~\ref{sec:appendix}.

Firstly, we note that we need $M \geq 4$ since there should be at least three different classes to distinguish three generations of leptons; The number of classes for non-invertible $Z_M$ is $p+1$ when $M= 2p$ or $2p+1$.

$M=4$ and $5$: In this case, we assign three classes $[g^{0,1,2}]$ for three generations of SM leptons $\{L(\bar{\ell})\}= \{L_e(\bar{e}), L_\mu(\bar\mu), L_\tau (\bar\tau) \}$ and three or two generations of $N$. Then we cannot forbid some elements of the $\bar{N}_\alpha L_j H$ operator while requiring non-zero elements of $\bar{L}_i \ell_j H$, leading to a type-I seesaw contribution. 

The above argument indicates that we need to distinguish the SM leptons and $N$ assigning different classes. Thus, we need $M \geq 8$
to assign at least two $N$ with different classes from those of the SM leptons.

$M=8$: For illustration, we first assign classes $\{[g^0], [g^1], [g^2]\}$ for both $\{L\}$ and $\{\bar{\ell} \}$, and $\{[g^3], [g^4] \}$ for two generations of $N$. 
The structure of Yukawa matrix for $\bar{L}N \eta$ depends on the choice of class for $\eta$. The possible structures are 
\begin{align}
y^\eta : \ \begin{pmatrix} 0 & 0 \\ * & 0 \\ 0 & * \end{pmatrix}_{(\eta:[g^2])}, \, \begin{pmatrix} * & 0 \\ 0 & * \\ * & 0 \end{pmatrix}_{(\eta:[g^3])}, \, \begin{pmatrix} 0 & * \\ * & 0 \\ 0 & 0 \end{pmatrix}_{(\eta:[g^4])},
\end{align}
where $"*"$ expresses a non-zero element, subscript indicates the choice of class for $\eta$ and we only list matrix with at least two non-zero elements.
We thus find there should be two different $\eta$ to fit neutrino data. 
Note also that we have non-zero elements for the operator $\bar{\ell} L \eta$ for above assignments of class for $\eta$ and stability of DM is not guaranteed.

We then consider all the possible assignment of classes for leptons where exchange of classes for $L(\bar{\ell})$ and $N$ from above case gives us different patterns. 
As a result, a viable case is found when classes are assigned as $\{[g^0], [g^4], [g^2]\}$ for both $\{L\}$ and $\{\bar{\ell} \}$, and $\{[g^3], [g^1] \}$ for two generations of $N$. 
In this case, we obtain structure of $y^\eta$ as 
\begin{align}
y^\eta : \ \begin{pmatrix} 0 & * \\ * & 0 \\ * & * \end{pmatrix}_{(\eta:[g^1])}, \, \begin{pmatrix} * & 0 \\ 0 & * \\ * & * \end{pmatrix}_{(\eta:[g^3])}.
\end{align}
Furthermore we can forbid $\bar{\ell} L \eta$ term since product of $\ell$ and $L$ give classes $[g^{k}]$ with $k$ being even number due to the assignment.
In the following section, we analyze the case as one of the benchmark models (BMs) where the assignment is summarized in Table.~\ref{tab:field_cont}. 
Note that we can also obtain other viable models by applying permutation within $\{[g^0], [g^4], [g^2]\}$ for $L(\bar{\ell})$.

$M=9$: The assignments of classes are the same as those of $M=8$, where we have classes $[g^0]$ to $g^{4}$. However, the structure of Yukawa coupling can be different. For example, when we assign $\{[g^0], [g^1], [g^2]\}$ for both $\{L\}$ and $\{\bar{\ell} \}$, and $\{[g^3], [g^4] \}$ for two generations of $N$, the structures of Yukawa coupling for $\bar{L} N \eta$ operator are summarized as
\begin{align}
y^\eta : \ \begin{pmatrix} 0 & 0 \\ * & 0 \\ 0 & * \end{pmatrix}_{(\eta:[g^2])}, \, \begin{pmatrix} * & 0 \\ 0 & 0 \\ 0 & * \end{pmatrix}_{(\eta:[g^3])}, \, \begin{pmatrix} 0 & * \\ * & * \\ * & 0 \end{pmatrix}_{(\eta:[g^4])},
\end{align}
where we only list matrix with at least two non-zero elements.
In this case, it is thus possible to realize neutrino mass matrix that can be fitted to observables with $\eta$ with $[g^4]$ class. 
However, we cannot forbid the term $\bar{\tau} L_\tau \eta$  and the stability of DM is not guaranteed. 
We also investigate all the possible assignments of classes but we can not find a viable model in this case. Note that we cannot guarantee stability of DM when we assign $\{[g^0], [g^4], [g^2]\}$ for both $\{L\}$ and $\{\bar{\ell} \}$, and $\{[g^3], [g^1] \}$ for two generations of $N$ in contrast to $M=8$ case. This is due to the property that singlet is a product of $[g^{k}]$ and $[g^{k'}]$ with both $k$ and $k'$ being even, allowing $[g^{k''}]$ with $k''$ being odd when $M$ is odd case.

$M=10$: For illustration, we first assign classes $\{[g^0], [g^1], [g^2]\}$ to both $\{L\}$ and $\{\bar{\ell} \}$, and $\{[g^3], [g^4], [g^5] \}$ for three generations of $N$. Then we obtain the structure of the Yukawa coupling for the operator $\bar{L} N \eta$ depending on the assignment class to $\eta$ such that
\begin{align}
y^\eta : \ \begin{pmatrix} * & 0 & 0 \\ 0 & *  & 0 \\ 0 & 0  & * \end{pmatrix}_{(\eta:[g^3])}, \ \begin{pmatrix} 0 & * & 0 \\ * & 0  & * \\ 0 & *  & 0 \end{pmatrix}_{(\eta:[g^4])}, \ \begin{pmatrix} 0 & 0 & * \\ 0 & *  & 0 \\ * & 0  & 0 \end{pmatrix}_{(\eta:[g^5])}, 
\end{align}
where we only list matrix with at least three non-zero elements. 
It would be possible to fit the neutrino observables by the second structure with $[g^4]$ for $\eta$, but we cannot forbid the $\bar{\tau} L_\tau \eta$ term that spoils the stability of DM.
Then we investigate all possible assignments as in the above cases. As a result, viable models are found as summarized in Table.~\ref{tab:viable-models} in the Appendix~\ref{sec:appendix2} that can fit neutrino data and guarantee stability of DM. 
The viable models are obtained when we assign classes with odd(even) $k$ and even(odd) $k'$ for $L(\bar{\ell})$ and $N$, respectively, as in the case of $M=8$.

$M=11$: The assignments of the classes are the same as those of $M=10$.
The structures of Yukawa coupling for the operator$\bar{L} N \eta$ are
\begin{align}
y^\eta : \ \begin{pmatrix} * & 0 & 0 \\ 0 & *  & 0 \\ 0 & 0  & * \end{pmatrix}_{(\eta:[g^3])}, \ \begin{pmatrix} 0 & * & 0 \\ * & 0  & * \\ 0 & 0  & * \end{pmatrix}_{(\eta:[g^4])}, \ \begin{pmatrix} 0 & 0 & * \\ 0 & *  & * \\ * & *  & 0 \end{pmatrix}_{(\eta:[g^5])}, 
\end{align}
where we only list matrix with at least three non-zero elements. 
Remarkably, the choice of $[g^5]$ for $\eta$ can provide the structure of the Yukawa coupling to fit neutrino observables and forbid all the terms in the operator $\bar{e} L \eta$.
We analyze this case as one of the BMs below where the assignment is summarized in Table.~\ref{tab:field_cont}~\footnote{It is also possible to take only two generations of $N$ where the neutrino mass matrix in the case is mentioned in ref.~\cite{Kobayashi:2025cwx}.}.
Note also that we can also obtain the other Yukawa structures permuting and/or swapping rows by changing the assignment of classes to leptons $\{L \}$ and $\{\bar{e}\}$.
In addition to the case, we find other viable models which are summarized in Table~\ref{tab:viable-models} in the Appendix~\ref{sec:appendix2}.
Note that, for viable models we listed,
there is an accidental $Z_2$ symmetry in the Lagrangian that coincides with the one in the original scotogenic model, at renormalizable level.
Thus, in such cases, the renormalizable Lagrangian is invariant under the $Z_2$ transformations; $N \to -N$ and $\eta \to - \eta$. 
This $Z_2$ symmetry will not be broken by radiative correction as in the original scotogenic model and stability of DM is guaranteed with the minimal field contents.

\begin{table}[t]
    \centering
    \begin{tabular}{cccc} \hline
    \noalign{\vskip 0mm}
      \multicolumn{4}{c}{BM1} \\
      \noalign{\vskip 0mm}
      \hline \hline
      \noalign{\vskip 1mm}
       Fields & $SU(2)_L$ & $U(1)_Y$ & $\mathbb{Z}_{8}$ \\ \noalign{\vskip 1mm} \hline
       \noalign{\vskip 1mm}
       L & $\pmb{2}$ & $-\frac{1}{2}$ & $\{[g^0],[g^4],[g^2]\}$ \\ \noalign{\vskip 1mm}
       $\bar{\ell}$ & $\pmb{1}$ & $~1$ & $\{[g^0],[g^4],[g^2]\}$ \\ \noalign{\vskip 1mm}
       $N$ & $\pmb{1}$ & $~0$ & $\{[g^3],[g^1]\}$ \\ \noalign{\vskip 1mm} \hline \noalign{\vskip 1mm}
       $H$ & $\pmb{2}$ & $~\frac{1}{2}$ & $[g^0]$ \\ \noalign{\vskip 1mm}
       $\eta$ & $\pmb{2}$ & $-\frac{1}{2}$ & $[g^1]$ \\ \noalign{\vskip 1mm} \hline \hline \noalign{\vskip 0mm}
       \multicolumn{4}{c}{BM2} \\ \noalign{\vskip 0mm} \hline \hline \noalign{\vskip 1mm}
        Fields & $SU(2)_L$ & $U(1)_Y$ & $\mathbb{Z}_{11}$ \\ \noalign{\vskip 1mm} \hline \noalign{\vskip 1mm}
        L & $\pmb{2}$ & $-\frac{1}{2}$ & $\{[g^0],[g^1],[g^2]\}$ \\ \noalign{\vskip 1mm}
        $\bar{\ell}$ & $\pmb{1}$ & $~1$ & $\{[g^0],[g^1],[g^2]\}$ \\ \noalign{\vskip 1mm}
        $N$ & $\pmb{1}$ & $~0$ & $\{[g^3],[g^4],[g^5]\}$ \\ \noalign{\vskip 1mm} \hline \noalign{\vskip 1mm}
        $H$ & $\pmb{2}$ & $~\frac{1}{2}$ & $[g^0]$ \\ \noalign{\vskip 1mm}
        $\eta$ & $\pmb{2}$ & $-\frac{1}{2}$ & $[g^5]$ \\ \noalign{\vskip 1mm} \hline \hline
    \end{tabular}
    \caption{Fields content of the benchmark models}
    \label{tab:field_cont}
\end{table}
%

\subsection{Charged Lepton Masses}
\label{sec:chrd_lep_mass}

In the model Yukawa coupling matrix for $\bar{L} \ell H$ is the diagonal matrix due to the symmetry.
Then charged lepton mass matrix is given by
\begin{equation}
M_e = \frac{v}{\sqrt{2}} \begin{pmatrix} y^\ell_{11} & 0 & 0 \\ 0 & y^\ell_{22} & 0 \\ 0 & 0 & y^\ell_{33} \end{pmatrix}
\end{equation}
where all elements are chosen to be real using the phase transformation of $e_{R_i}$. 
Then masses of charged lepton are simply given by $\{m_e, m_\mu, m_\tau\} = \{v \, y^\ell_{11}/\sqrt{2}, v \, y^\ell_{22}/\sqrt{2}, v \, y^\ell_{33}/\sqrt{2} \}$ and all charged leptons are mass eigenstates without rotation.

%
\subsection{Neutrino Masses}
\label{sec:mnu}

Here we formulate neutrino masses for the model at hand.
Firstly, for benchmark model 1[2], the mass matrix $M_N$ is diagonal written by

\begin{align}
& M_N =  \begin{pmatrix} M_1 & 0  \\ 0 & M_2  \end{pmatrix} \quad \text{(BM1)}, \\  
& M_N = \begin{pmatrix} M_1 & 0 & 0 \\ 0 & M_2 & 0 \\ 0 & 0 & M_3 \end{pmatrix} \quad \text{(BM2)},
\end{align}

where the elements can be real using the phase transformation of $N_\alpha \, (\alpha=1,2 [1,2,3])$.
Thus, $N$s are mass eigenstates without rotation.  

The active neutrino mass matrix is generated at the one-loop level as in the original scotogenic model~\cite{Ma:2006km}.
The Yukawa coupling $y^\eta$ is written as
\begin{align}
y^\eta &= \begin{pmatrix} 0 & y^\eta_{12}  \\ y^\eta_{21}  & 0 \\ y^\eta_{31} e^{i \alpha}  & y^\eta_{32}  \end{pmatrix} \quad \text{(BM1)}, \label{eq:Yukawa-structure1} \\ 
y^\eta &= \begin{pmatrix} 0 & 0 & y^\eta_{13} \\ 0 & y^\eta_{22} & y^\eta_{23} e^{i \alpha} \\ y^\eta_{31} & y^\eta_{32} e^{i \beta} & 0   \end{pmatrix} \quad \text{(BM2)}, \label{eq:Yukawa-structure2}
\end{align}
where $y^\eta_{ij}$ is real without loss of generality. We have two remaining phases $\{\alpha, \beta \}$ as we can eliminate three phases by phase transformation of $L_i$.
Calculating the one-loop diagram, the neutrino mass matrix is approximately obtained as 
\begin{subequations}
\label{eq:neutrino_masses}
\begin{align}
\label{eq:neutrino_masses_a}
& (m_\nu)_{ij} \simeq \sum_{\alpha=1-3} \frac{y^\eta_{i \alpha} M_\alpha F_\alpha (y^{\eta})^T_{\alpha j}}{(4 \pi)^2},  \\
\label{eq:neutrino_masses_b}
& F_\alpha =  \frac{m_R^2}{m_R^2 - M^2_\alpha} \ln \left[ \frac{m_R^2}{M^2_\alpha}\right] - \frac{m_I^2}{m_I^2 - M^2_\alpha} \ln \left[ \frac{m_I^2}{M^2_\alpha}\right], 
\end{align}
\end{subequations}
where the charged lepton mass is ignored.

The structure of the neutrino mass matrix is determined by that of the Yukawa matrix $y^\eta$ since the Majorana mass matrix for $N$ is diagonal. We then obtain a one-zero texture of the neutrino mass matrix. The location of the vanishing element is shifted by changing the assignment of class to leptons $L(\bar{e})$ as summarized in Table~\ref{tab:structure} for $M=8$ and $M=11$ where the assignments for $N$ are the same as BM1 and BM2; in principle, any one-zero texture can fit the neutrino data~\cite{Lashin:2011dn}, but it is not trivial if the number of free parameters is less than general one-zero texture matrices.

\begin{table}[t]
    \centering
    \begin{tabular}{cccc}
        \hline 
        \noalign{\vskip 2mm}
        & \multicolumn{3}{c}{Assignment of classes for $\{L(\bar{\ell})\}$} \\ 
        \noalign{\vskip 2mm}
        $M=8$ & $\{[g^0],[g^2],[g^4]\}$  & $\{[g^2],[g^4],[g^0]\}$  & $\{[g^0],[g^4],[g^2]\}$  \\
        \noalign{\vskip 2mm}
        $M=11$ &  $\{[g^0],[g^1],[g^2]\}$ & $\{[g^1],[g^0],[g^2]\}$ &  $\{[g^0],[g^2],[g^1]\}$ \\ 
        \noalign{\vskip 2mm}
        \hline
        \noalign{\vskip 2mm}
         $m_\nu$ & $\begin{pmatrix} * & * & 0 \\ * & * & * \\ 0 & * & * \end{pmatrix}$  & $\begin{pmatrix} * & * & * \\ * & * & 0 \\ * & 0 & * \end{pmatrix}$ & $\begin{pmatrix} * & 0 & * \\ 0 & * & * \\ * & * & * \end{pmatrix}$ \\ 
         \noalign{\vskip 2mm}
         \hline 
         \noalign{\vskip 2mm}
         cLFV & $\mu \to e \gamma$ & $\mu \to e \gamma$ & $\tau \to e \gamma$ \\ 
         \noalign{\vskip 2mm} 
          & $\tau \to \mu \gamma$ & $\tau \to e \gamma$ & $\tau \to \mu \gamma$ \\
         \noalign{\vskip 2mm} 
         \hline
    \end{tabular}
    \caption{Relation among assignment of classes to $\{L(\bar{\ell})\} = \{L_e(\bar{e}), L_\mu (\bar{\mu}), L_\tau (\bar{\tau}) \}$, a structure of neutrino mass, and cLFV modes with non-zero BRs at one loop level in $M=8$ and $M=11$ cases. The assignments for $N$ are the same as BM1 and BM2.}
    \label{tab:structure}
\end{table}

The neutrino mass matrix can be diagonalized by the unitary matrix $U_\nu$ as $m_\nu^d = U_\nu^T m_\nu U_\nu$ where $m_\nu^d = {\rm diag}(m_1, m_2, m_3)$.
The unitary matrix $U_\nu$ is identified as the Pontecorvo-Maki-Nakagawa-Sakata (PMNS) mixing matrix, since the charged lepton mass matrix is diagonal.
In our analysis, we choose the notation of the Majorana phases such as $[1,e^{i\alpha_{21}/2},e^{i\alpha_{31}/2}]$. 
Then we numerically extract the neutrino mixing angles and phases from the PMNS matrix. In addition, effective mass for neutrinoless double beta decay is estimated by 
\begin{align}
|m_{ee}|=\left| m_{1} c^2_{12} c^2_{13}+m_{2} s^2_{12} c^2_{13}e^{i\alpha_{21}}
+m_{3} s^2_{13}e^{i(\alpha_{31}-2\delta_{\mathrm{CP}})} \right|,
\end{align}
where $\delta_{\mathrm{CP}}$ is the Dirac CP phase in PMNS matrix.
Currently, the upper bound is given by $| m_{ee}| <(28-122)$ meV at 90 \% confidence level~\cite{KamLAND-Zen:2024eml} where the range of the upper limit is due to the adopted model of the nuclear mass matrix elements.

In the numerical analysis below, the neutrino mass matrix is rewritten by
\begin{equation}
m_\nu(m_{\eta_R}, m_{\eta_I}, M_{\alpha}) = M_1 \tilde{m}_\nu (\tilde{m}_{\eta_R}, \tilde{m}_{\eta_I}, \tilde{M}_{\alpha}),
\end{equation}
where $\tilde{m}_\nu$ is the dimensionless matrix and we define dimensionless parameters $\tilde{m}_{\eta_R,\eta_I} \equiv m_{\eta_R,\eta_I}/M_1$ and $\tilde{M}_{\alpha} \equiv M_{\alpha}/M_1$.
We denote the eigenvalues of $\tilde{m}_\nu$ by $\{\tilde{m}_1,\tilde{m}_2,\tilde{m}_3 \}$.
Then, overall factor $M_1$ can be estimated from the following relation;
\begin{align}
{\rm(NH):} \ M_1^2 = \frac{\Delta m_{\rm atm}^2}{\tilde{m}_3^2 - \tilde{m}_1^2}, \quad
{\rm(IH):} \ M_1^2 = \frac{\Delta m_{\rm atm}^2}{\tilde{m}_2^2 - \tilde{m}_3^2}, 
\end{align}
where $\Delta m_{\rm atm}^2$ is the atmospheric neutrino mass square difference, and NH and IH, respectively, stand for the normal and inverted hierarchy.

\subsection{Charged lepton flavor violation}
\label{sec:fcnc}
Charged lepton flavor violation (cLFV) processes are induced at one-loop level through Yukawa interaction. 
Calculating one-loop diagrams, we obtain the BRs such that~\cite{Baek:2016kud}
\begin{subequations}
\label{eq:clfv}
\begin{align}
\label{eq:br_li_ljgamma}
&{\rm BR}(\ell_i\to\ell_j\gamma) \nonumber \\ 
& \approx
\frac{48\pi^3\alpha_{em}C_{ij}}{G_F^2 (4\pi)^4}
\left|\sum_{\alpha=1-3}y^\eta_{j\alpha} y^{\eta \dag}_{\alpha i} F(M_{\alpha},m_{\eta^\pm})\right|^2, \\
\label{eq:f_func_clfv}
&F(m_a,m_b) \nonumber \\ & \approx \frac{2 m^6_a+3m^4_am^2_b-6m^2_am^4_b+m^6_b+12m^4_am^2_b\ln\left(\frac{m_b}{m_a}\right)}{12(m^2_a-m^2_b)^4},
\end{align}
\end{subequations}
where $C_{21}=1$, $C_{31}=0.1784$, $C_{32}=0.1736$, $\alpha_{em}(m_Z)=1/128.9$ is the fine structure constant and $G_F=1.166\times10^{-5}$ GeV$^{-2}$ is the Fermi constant.
Here $C_{ij}$ is defined by $C_{ij} \equiv \Gamma(\ell_i \to \ell_j \overline{\nu_j} \nu_i)/\Gamma_{\ell_i}^{\rm total}$ where the numerator is partial decay width and denominator is the total decay width of $\ell_i$ in the SM.
We note that possible cLFV mode is related to the structure of the neutrino mass through $y^\eta$ and we have summarized modes with non-zero branching ratio(BR) in Table~\ref{tab:structure}.
In the numerical analysis below we impose current experimental upper bounds of the BRs:
by~\cite{MEG:2016leq, BaBar:2009hkt,Renga:2018fpd}
\begin{subequations}
\label{eq:lfvs-cond}
\begin{align}
& {\rm BR}(\mu\to e\gamma)\lesssim 3.1\times10^{-13}, \\
& {\rm BR}(\tau\to e\gamma)\lesssim 3.3\times10^{-8}, \\
& {\rm BR}(\tau\to\mu\gamma)\lesssim 4.4\times10^{-8}.
\end{align}
\end{subequations}

The same one-loop diagram inducing $\mu \to e \gamma$ provides dipole operator which also enables the $\mu$-$e$ conversion process. The current limit on the ratio of $\mu$-$e$ conversion is ${\rm CR}(\mu \to e)_{Au} \leq 7.0 \times 10^{-13}$~\cite{SINDRUMII:2006dvw}.
Furthermore, three body CLFV decays $\ell^\mp_i \to \ell^\mp_j \ell^\mp_{k} \ell^\pm_{l}$ can also be induced via the Yukawa interactions. Several one-loop diagrams can contribute to the processes such as $\gamma$-penguins, $Z$-penguins, Higgs-penguins and Box diagrams. The flavor structure of penguin diagrams are the same as $\ell_i \to \ell_j \gamma$ while the box diagrams depend on the structure of Yukawa coupling $y^\eta$. The constraints on the three-body decays are given by~\cite{SINDRUM:1987nra,Hayasaka:2010np,Belle-II:2024sce} 
\begin{subequations}
\label{eq:lfvs-cond}
\begin{align}
& {\rm BR}(\mu \to 3 e) \leq 1.0 \times 10^{-12}, \\
& {\rm BR}(\tau \to 3e) \leq 2.7 \times 10^{-8}, \\
& {\rm BR}(\tau \to 3 \mu) \leq 1.9 \times 10^{-8}. 
\end{align}
\end{subequations}
In the benchmark models, the strongest constraint is imposed by the upper limit of ${\rm BR}(\mu \to e \gamma)$; in fact, unless the mass differences among $N_i$, $\eta_{R(I)}$ and $\eta^{\pm}$ are very hierarchical, ratios of $\mu \to e$ conversion and $\mu \to 3e$ are smaller than that of $\mu \to e\gamma$~\cite{Toma:2013zsa}. 
We thus focus on $\ell_i \to \ell_j \gamma$ constraints in the numerical analysis below.

\section{Numerical Analysis}
\label{sec:num_ana}
In this section, we carry out numerical analysis for neutrino mass matrix and cLFV. For illustration, we discuss the structure of $y^\eta$ in Eq.~\eqref{eq:Yukawa-structure1} and \eqref{eq:Yukawa-structure2} that induces one-zero textures of $m_\nu$.

For BM1(BM2), we randomly scan our non-zero Yukawa couplings in Eq.~\eqref{eq:Yukawa-structure1} and \eqref{eq:Yukawa-structure2}  with one(two) phases $\{y^\eta_{ij}, \alpha (\alpha, \beta)  \}$ and mass related parameters $\{\tilde{M}_{2(2,3)}, \tilde m_{\eta_R}, \Delta m^2 \}$  to fit the neutrino data within the range of 
\begin{align}
& \{ \tilde{M}_{2(2,3)}, \tilde{m}_{\eta_R} \} \in [1, 10], \
\Delta m^2 \in [10^{-6}, 0.1], \\
& \{y^\eta_{ij} \} \in [10^{-8}, 1], \ \{\alpha (\alpha, \beta) \} \in [-\pi, \pi]. \nonumber 
\end{align}
The parameter $M_1$ is fixed to fit $\Delta m_{\rm atm}^2$ where we scan the value within 3$\sigma$ confidence level.
Then we search for the allowed parameter sets that can fit the neutrino data from Nufit 6.0~\cite{Esteban:2024eli}.
The value of $\Delta \chi^2$ is estimated using the formula of
\begin{equation}
\Delta \chi^2 = \sum_{i}  \left( \frac{O_i^{\rm obs} - O_i^{\rm th}}{\delta O_i^{\rm exp}} \right)^2, \label{eq:chi-square}
\end{equation}
where $O_i^{\rm obs (th)}$ is the observed (theoretically) obtained value of corresponding observables and $\delta O_i^{\rm exp}$ represents the experimental error at $1\sigma$ level.
Notice that, we calculate $\Delta \chi^2$ with $\{\sin^2 \theta_{12}, \sin^2 \theta_{13}, \Delta m^2_{\rm atm}, \Delta m^2_{sol} \}$ and $\sin^2 \theta_{23}$ is just required to be within 3$\sigma$ range by Nufit 6.0 since the experimental error is deviated from Gaussian distribution.

Firstly, we find that all the neutrino data can be fitted for IH case only in BM1. On the other hand, we can fit all the data for both the NH and the IH cases in BM2. Since $M=8$ cases have less free parameters it is more difficult to fit the data compared with $M=10$ and $M=11$ models. In the following, we show our results for our allowed parameter space in BM1 (IH case) and BM2 (NH and IH cases).
%
%

\begin{figure*}[tb]
\begin{center}
\includegraphics[width=0.31\textwidth]{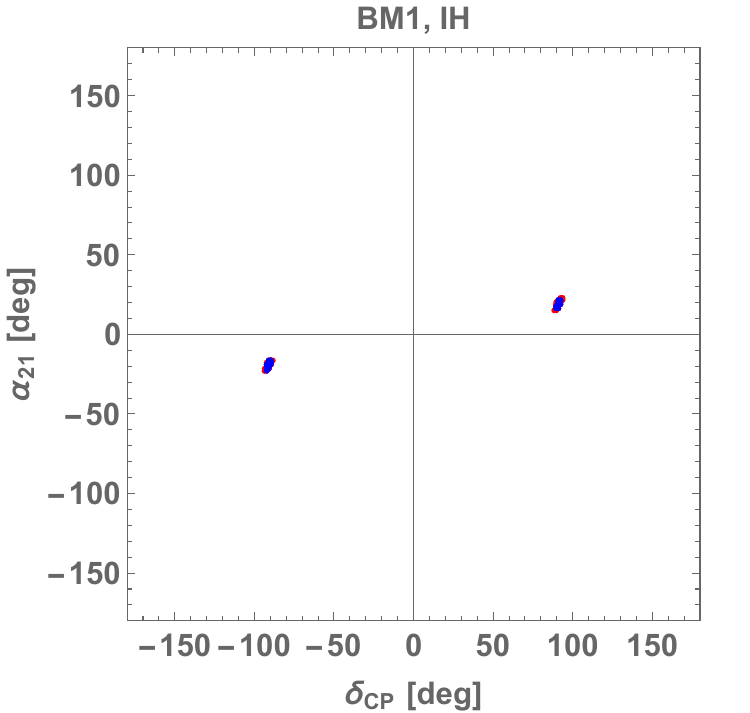} \quad
\vspace{2mm}
\includegraphics[width=0.31\textwidth]{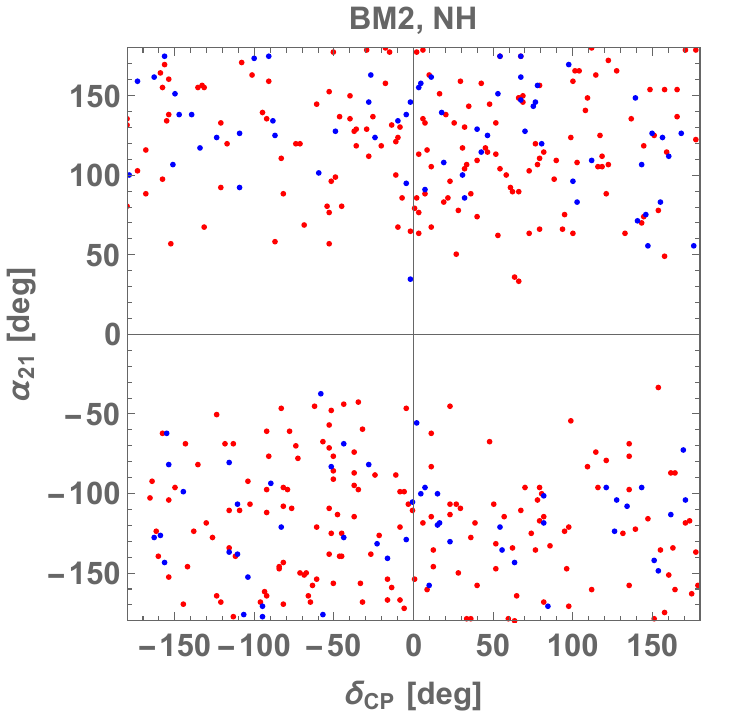} \quad
\vspace{2mm}
\includegraphics[width=0.31\textwidth]{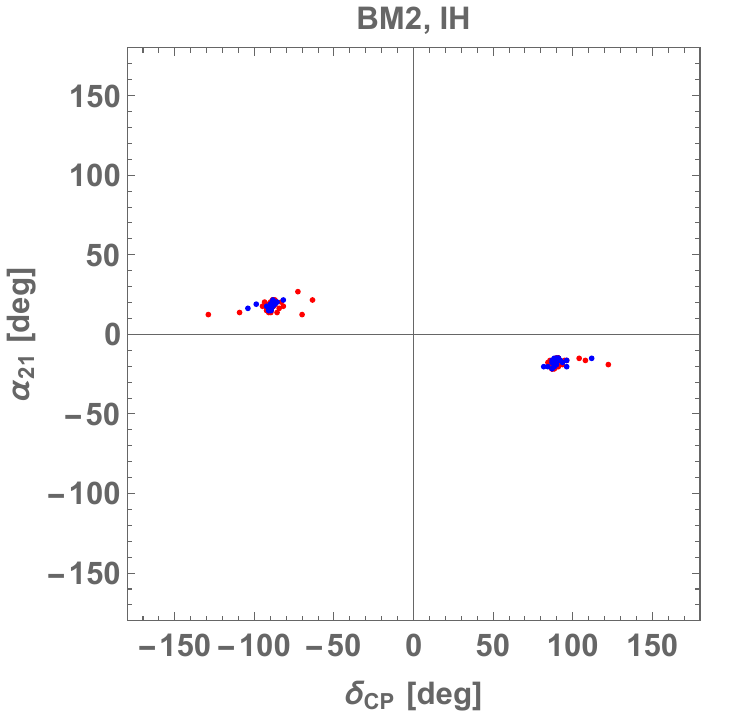} 
\caption{Values of $\delta_{\rm CP}$ and $\alpha_{21}$ estimated from allowed parameters where red and blue color points give $\Delta\chi^2$ values corresponding to the confidence level within $5 \sigma $ to $3 \sigma$ and within $3 \sigma$ level (the color legend is common in all the figures).}
  \label{fig:angle1}
\end{center}\end{figure*}

Fig.~\ref{fig:angle1} shows the values of $\delta_{\rm CP}$ and $\alpha_{21}$ by our allowed parameter sets where red and blue color points give $\Delta\chi^2$ values corresponding to the confidence level within $5 \sigma $ to $3 \sigma$ and within $3 \sigma$ level (the color legend is common in all the figures). For BM1, we find that the predicted region is very narrow, where $\delta_{\mathrm{CP}} \sim \pm 90$[deg] and $\alpha_{21} \sim \pm 20$[deg].
These phases are constrained via Eq.~\eqref{eq:oneZero-const6}.
For BM2  the left(right) plot corresponds to the NH(IH) case. 
For the NH case, we find that $\delta_{\rm CP}$ can be any value while $|\alpha_{21}|$ value is preferred to be within $[50, 180]$ [deg], and there is no clear correlation between them.
For the IH case, the value $\delta_{\rm CP}$ is preferred to be around $\pm 90$ [deg] and the value $\alpha_{21}$ is also preferred to be around $[10, 30]$ [deg] without a clear correlation between them.
The phases are constrained via Eq.~\eqref{eq:oneZero-const4}. For NH case, all three terms in the LHS of the equation cannot be ignored and we have less constraint on the phases. For IH case, first two terms in the LHS rather dominate and the phases are more constrained.

\begin{figure}[tb]
\begin{center}
\includegraphics[width=62.0mm]{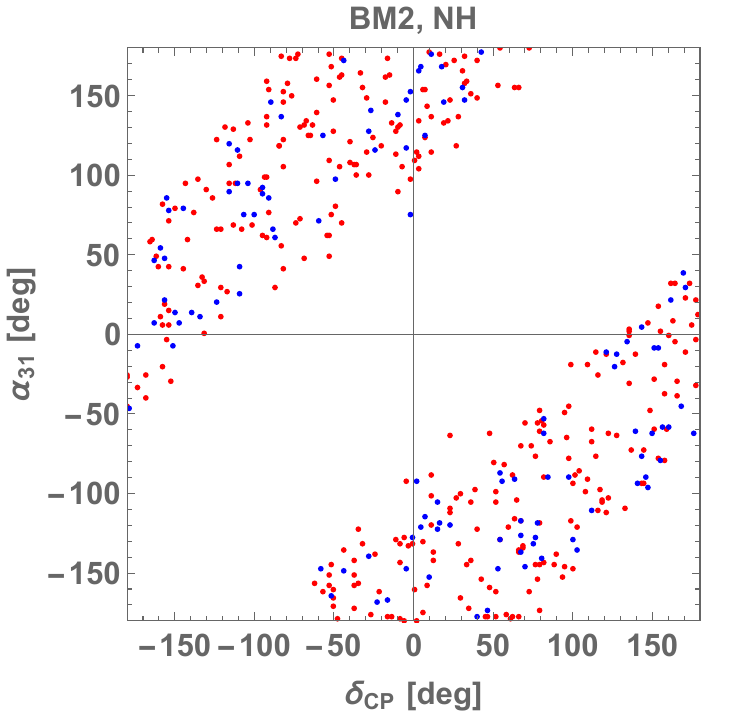} \\
\includegraphics[width=62.0mm]{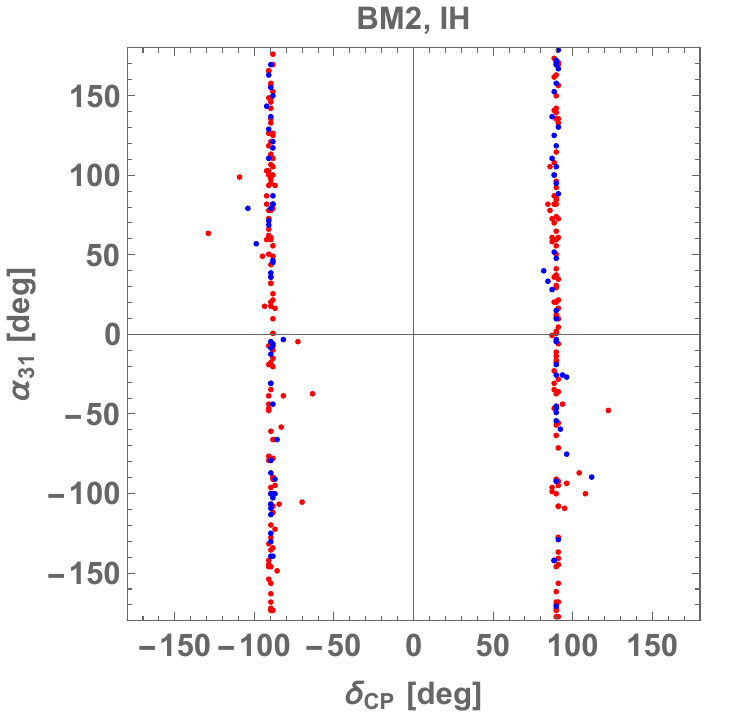} 
\caption{Values of $\delta_{\rm CP}$ and $\alpha_{31}$ estimated from allowed parameters.}
  \label{fig:angle2}
\end{center}\end{figure}

Fig.~\ref{fig:angle2} shows the values of $\delta_{\rm CP}$ and $\alpha_{31}$ by our allowed parameter sets where the left(right) plot corresponds to the NH(IH) case.
For BM1, the $\alpha_{31}$ is not physical since we only have two non-zero neutrino masses.
For BM2, any value of $\alpha_{31}$ can also be realized and there is a clear correlation between $\delta_{\rm CP}$ in the NH case while no correlation is found in the IH case.
The $\alpha_{31}$ is also constrained via Eq.~\eqref{eq:oneZero-const4}. For NH case, all three terms in the LHS of the equation cannot be ignored and the third term is little larger than the others that provides little correlation between $\alpha_{31}$ and $\delta_{\rm CP}$. For IH case, first two terms in the LHS rather dominate and $\alpha_{31}$ can be any value.

\begin{figure*}[tb]
\begin{center}
\includegraphics[width=0.31\textwidth]{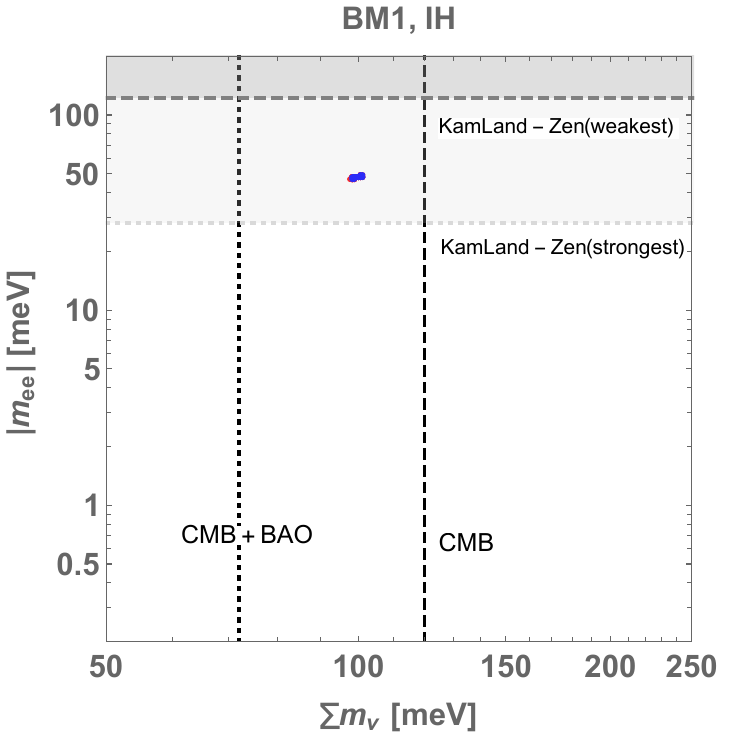} \quad
\includegraphics[width=0.31\textwidth]{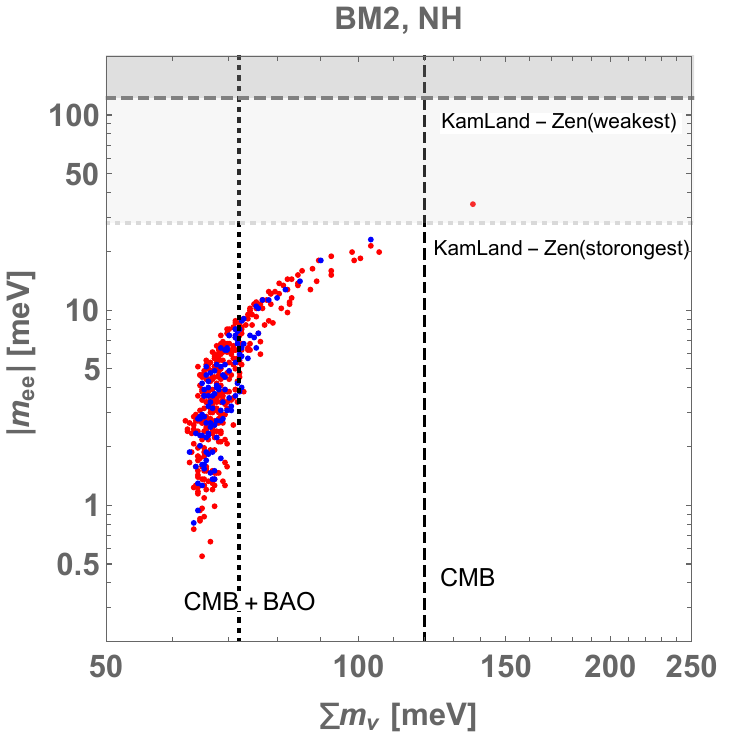} \quad
\includegraphics[width=0.31\textwidth]{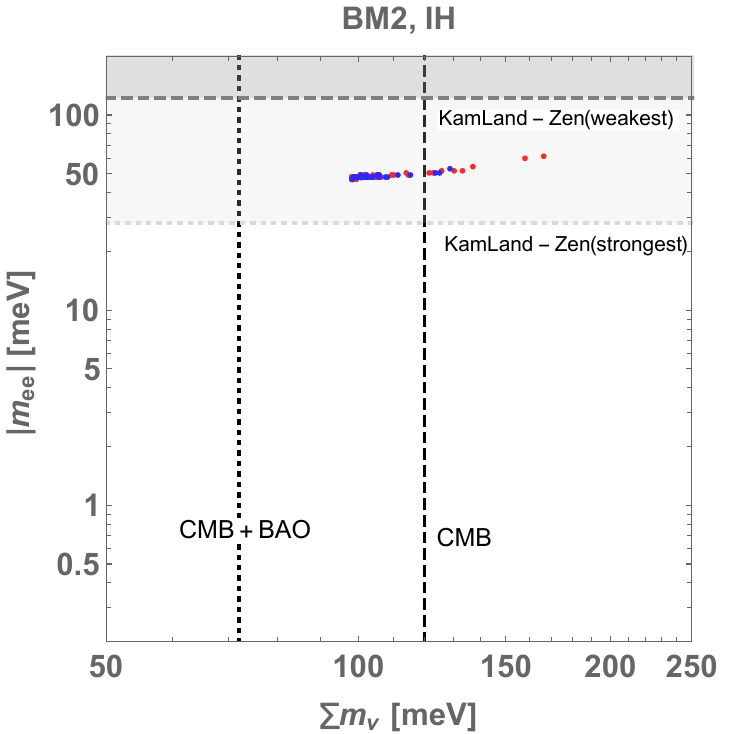}
\caption{Values of $\sum m_\nu$ and $|m_{ee}|$ estimated from allowed parameters.}
  \label{fig:mass1}
\end{center}
\end{figure*}

Fig.~\ref{fig:mass1} shows the predicted values of $\sum m_\nu$ and $m_{ee}$ from our allowed parameter sets . The dashed and dotted vertical lines respectively indicate the cosmological bound on $\sum m_\nu$ by the Planck CMB data~\cite{Planck:2018vyg} and by combination of the CMB and the DESI BAO data~\cite{DESI:2024mwx}, and the dashed and dotted horizontal lines respectively represent the weakest and strongest constraint from current KamLand-Zen~\cite{KamLAND-Zen:2024eml}.
For BM1, we find that very limited region in the IH where $\sum m_\nu \sim 100$ meV and $|m_{ee}| \sim 50$ meV. 
For BM2, we have points that satisfy both the cosmological bounds on $\sum m_\nu$ for the NH case, while there are points that only satisfy the CMB limit for the IH case.
In addition, the points in the NH case are below the current $0\nu \beta \beta$ bound while the points in the IH case are between the strongest and weakest upper limits on $|m_{ee}|$.
Thus, the IH case could be tested in the near future.

\begin{figure}[tb]
    \begin{center}
        \includegraphics[width=62.0mm]{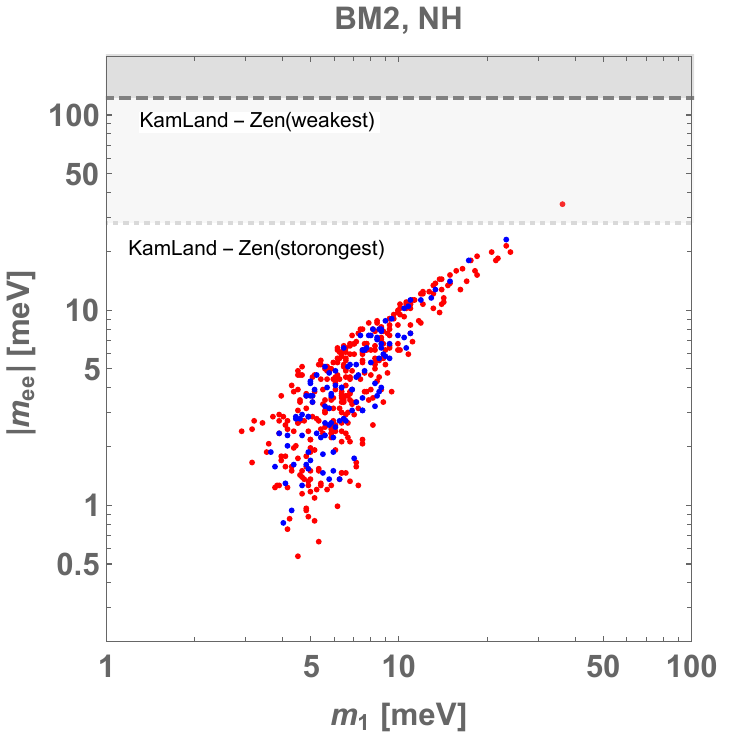} \\
        \includegraphics[width=62.0mm]{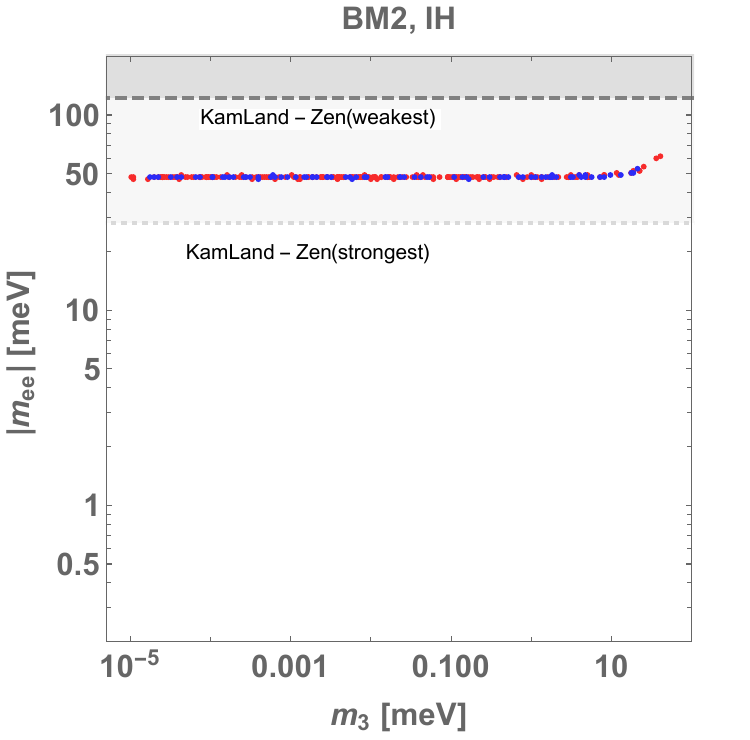}
        \caption{Values of $m_1(m_3)$ and $m_{ee}$ for NH(IH) estimated from allowed parameters.}
        \label{fig:mass3}
    \end{center}
\end{figure}

Finally, Fig.~\ref{fig:mass3} shows the predicted values in the $\{m_1, m_{ee} \}$ plane for BM2. The values of the lightest neutrino mass are within $[1, 40] ([10^{-5}, 20])$ meV for the NH(IH) case.
Note that we do not show a plot for BM1 since the lightest neutrino mass is always zero since we have only two generation of $N$ making the $m_\nu$ rank 2 matrix.

Before closing the section, we comment on the effects of quantum correction, where one-loop vertex correction would induce elements of Yukawa couplings which are zero at tree level. 
Let us consider the most hierarchical case that is the IH case in BM2 where ${\rm Max}[y^\eta]/{\rm Min[y^\eta]} = 10^3$ and the largest(smallest) Yukawa coupling is ${\rm Max}[y^\eta]({\rm Min}[y^\eta]) \lesssim 0.01(10^{-5})$. We could have quantum correction contributing elements that are zero at tree level. However, it will be sufficiently small since the one-loop vertex correction should give $(y^\eta)^3/(4\pi)^2$ factor; applying ${\rm Max[y^\eta]}$ it is $\mathcal{O}(10^{-8})$.
Therefore, it is safe to ignore the quantum correction in analyzing neutrino observables.

%
%

%
\section{Summary and Discussion}
\label{sec:discussion}

In this work, we have discussed scotogenic models with non-invertible $Z_M$ symmetry.
The three generations of standard model leptons and singlet Majorana fermions are distinguished by the assignment of conjugacy classes under the $Z_M$ symmetry.
We have then searched for viable models that realize the generation of scotogenic neutrino mass and the stability of dark matter.
We find $M=8$ to provide a viable model and analyze one model as our benchmark where neutrino data can be fitted in only IH case due to the lack of free parameters. 
On the other hand, our benchmark model with $M=11$ can fit the data in which six classes are assigned to the three generations of standard model leptons and singlet fermions. 
Note that other models with $M=10$ and $M=11$ can also fit the data since they have the same number of free parameters and realize one-zero textures.

The neutrino mass matrix in the minimal case has a one-zero texture where the position of vanishing element depends on the assignment of classes. 
For illustration, we have performed a numerical analysis for the case in which the element $31(13)$ is zero.
It has been then found that we have some correlation and preferences for the Dirac CP phase and Majorana phases that depend on the hierarchy of neutrino masses.
Then, the phenomenological difference between vanilla scotogenic model and non-invertible scenario reveals itself at the flavor level of the Lagrangian. In particular, the difference becomes apparent in the lepton mass matrices and lepton mixing patterns. In addition, we find some charged lepton flavor violating processes are absent depending on the assignment under the symmetry, which is also related to the one-zero structure of the neutrino mass matrix.

Thus we can have a scotogenic model with more predictability by modifying the original $Z_2$ symmetry group to no-group $Z_M$ symmetry. 

\acknowledgments
The work was supported by the National Natural Science Fund of China Grant No.~12350410373 (O.~P.) and by the Fundamental Research Funds for the Central Universities (T.~N.). 

The order of the authors' names is alphabetical.
%
%
\appendix

\section{Selection rules in non-invertible $Z_M$ symmetry} \label{sec:appendix}

A non-invertible $Z_M$ symmetry can be realized considering $T_2/Z_2$ orbifold with gauged $Z_2$ as an extra dimension that comes from the context of extra dimensional D-brane models of type IIB superstring theory with magnetic flux.  For more details, we can refer to refs.~\cite{Kobayashi:2024yqq,Kobayashi:2024cvp,Kobayashi:2025znw} for theoretical background. Here we summarized basic selection rules.
First, we consider $Z_M$ symmetries with generator $g$. 
Then a non-invertible symmetry can be constructed using the automorphism 
\begin{equation}
e g^{-1} e = g, \quad r g r^{-1} = g^{-1}.
\end{equation}
We can define the following class using the automorphism such that 
\begin{equation}
[g^k] = \{ h g^k h^{-1} \ | \ h = e, r  \} \quad (k = 0, 1, \cdots, M-1),
\end{equation}
where $\{e,r\}$ is associated with gauged $Z_2$ symmetry on orbifold.
In this definition, there are $p+1$ kinds of classes for $M = 2p$ and $M=2p +1$ cases; note that $[g^{k}] = [g^{M-k}]$. 

In the no group $Z_M$ symmetry scheme, we assign a class $[g^k]$ to a field.
The product of two classes is given by
\begin{equation}
[g^k] [g^{k'}] = [g^{k+k'}] + [g^{M-k+k'}]. \label{eq:products}
\end{equation}
Then, for example, when two fields $\varphi_k$ and $\varphi_{k'}$ belong to $[g^k]$ and $[g^{k'}]$ the term $\varphi_k \varphi_{k'}$ is allowed by the symmetry 
if the right hand sides of Eq.~\eqref{eq:products} include $[g^0]$.
The rule is easily applied to terms with more fields. 
For three fields $\varphi_k$, $\varphi_{k'}$ and $\varphi_{k''}$ with $\varphi_{k''}$ belonging to $[g^{k''}]$, the term $\varphi_k \varphi_{k'} \varphi_{k''}$ is allowed if the products $[g^{k}] [g^{k'}] [g^{k''}]$ give us $[g^0]$ under the rule of Eq.~\eqref{eq:products}.

\section{Summary of viable models} \label{sec:appendix2}
In this appendix we summarize the viable models and comment on phenomenology. The representative ones are summarized in Table~\ref{tab:viable-models}.
In addition to the structures of $y^\eta$ and $m_\nu$ in the table, we can obtain other structures by taking a permutation of assignments of classes to leptons. 
For example, we consider model class (1) with $Z_8$. Starting from the assignment in the table, we can obtain
\begin{align}
\{[g^0], [g^2], [g^4]\}_{L(\bar e)} : y^\eta = \begin{pmatrix} 0 & * \\ * & * \\ * & 0 \end{pmatrix}, \ m_\nu = \begin{pmatrix} * & * & 0 \\ * & * & * \\ 0 & * & * \end{pmatrix}, \nonumber \\
\{[g^2], [g^4], [g^0]\}_{L(\bar e)} : y^\eta = \begin{pmatrix} * & * \\ * & 0 \\ 0 & * \end{pmatrix}, \ m_\nu = \begin{pmatrix} * & * & * \\ * & * & 0 \\ * & 0 & * \end{pmatrix}, 
\end{align}
where assignments for $N$ and $\eta$ are not changed.
Thus the possible one-zero structures of $m_\nu$ are 
\begin{align}
\label{eq:oneZero}
\begin{pmatrix} * & 0 & * \\ 0 & * & * \\ * & * & * \end{pmatrix}, \quad
\begin{pmatrix} * & * & 0 \\ * & * & * \\ 0 & * & * \end{pmatrix}, \quad
\begin{pmatrix} * & * & * \\ * & * & 0 \\ * & 0 & * \end{pmatrix}.
\end{align}
We can realize these one-zero texture matrices for all model classes in Table~\ref{tab:viable-models} by taking a permutation of class assignments.
These textures provide some relations among neutrino observables such as mixing angles, masses, and phases as we show in the next appendix.

In addition, we have a clear correlation between neutrino mass matrix and squared amplitude for $\ell_i \to \ell_j \gamma$ process in Eq.~\eqref{eq:clfv}. 
For different textures, the $\ell_i \to \ell_j \gamma$ modes allowed at one-loop level are as follows
\begin{align}
& (m_\nu)_{12} =0 :  \tau \to e \gamma, \quad \tau \to \mu \gamma, \nonumber  \\
& (m_\nu)_{13} =0 :  \mu \to e \gamma, \quad \tau \to \mu \gamma,   \\
& (m_\nu)_{23} =0 :  \mu \to e \gamma, \quad \tau \to e \gamma. \nonumber
\end{align}
On the other hand, the other modes are forbidden at one-loop level. Similarly The $\mu$-$e$ conversion process is forbidden at one-loop level for $(m_\nu)_{12} =0$ case. 

Relations among the three body cLFV processes such as $\mu \to 3e$ and $m_\nu$ are not trivial. Since box diagrams contributing to three body decays are proportional to products of four Yukawa coupling $y^\eta$, the decay amplitudes depend on the structure of $y^\eta$. 
In addition, the structure of the Yukawa coupling affects collider physics since it determines a decay pattern of $N_i$ or components of $\eta$ depending on which one is the DM. For example, when $N_1$ is DM $\eta^\pm$ decays into $\ell^\pm N_1$ and lepton flavor dependence of the final state is given by $y^\eta$. Thus we can test the structure of $y^\eta$ by searching for signals of $\eta^\pm$ produced via electroweak interactions at collider experiments.
However it is beyond the scope of this work and further exploration is left for future works.

\begin{table*}[t]
    \centering
    \begin{tabular}{ccccccc}
     \hline \hline
        \noalign{\vskip 1mm}
         \hspace{0.5cm} Model class \hspace{0.5cm} & \hspace{0.5cm} $Z_M$ \hspace{0.5cm} & $L(\bar{e})$ & $N$ & \hspace{0.5cm} $\eta$ \hspace{0.5cm} & \hspace{1.0cm} $y^\eta$ \hspace{1.0cm} & \hspace{1.0cm} $m_\nu$ \hspace{1.0cm} \\ \noalign{\vskip 1mm} \hline \noalign{\vskip 1mm}
         (1) & $Z_8$ & $\{[g^0], [g^4], [g^2]\}$ & $\{[g^3], [g^1] \}$ & $[g^1]$ & $\begin{pmatrix} 0 & * \\ * & 0 \\ * & * \end{pmatrix}$ & $\begin{pmatrix} * & 0 & * \\ 0 & * & * \\ * & * & * \end{pmatrix}$ \\ \noalign{\vskip 1mm} \hline \noalign{\vskip 1mm}
         (2) & $Z_8$ & $\{[g^0], [g^4], [g^2]\}$ & $\{[g^3], [g^1] \}$ & $[g^3]$ & $\begin{pmatrix} * & 0 \\ 0 & * \\ * & * \end{pmatrix}$ & $\begin{pmatrix} * & 0 & * \\ 0 & * & * \\ * & * & * \end{pmatrix}$ \\ \noalign{\vskip 1mm} \hline \noalign{\vskip 1mm}
         (3) & $Z_{10}$ & $\{[g^0], [g^4], [g^2]\}$ & $\{[g^3], [g^1], [g^5] \}$ & $[g^3]$ & $\begin{pmatrix} * & 0 & 0 \\ * & * & 0 \\ 0 & * & * \end{pmatrix}$ & $\begin{pmatrix} * & * & 0 \\ * & * & * \\ 0 & * & * \end{pmatrix}$ \\ \noalign{\vskip 1mm} \hline \noalign{\vskip 1mm}
         (4) & $Z_{10}$ & $\{[g^3], [g^1], [g^5]\}$ & $\{[g^0], [g^4], [g^2] \}$ & $[g^1]$ & $\begin{pmatrix} 0 & * & * \\ * & 0 & * \\ 0 & * & 0 \end{pmatrix}$ & $\begin{pmatrix} * & * & * \\ * & * & 0 \\ * & 0 & * \end{pmatrix}$ \\ \noalign{\vskip 1mm} \hline \noalign{\vskip 1mm}
         (5) & $Z_{11}$ & $\{[g^0], [g^1], [g^2]\}$ & $\{[g^3], [g^4], [g^5] \}$ & $[g^5]$ & $\begin{pmatrix} 0 & 0 & * \\ 0 & * & * \\ * & * & 0 \end{pmatrix}$ & $\begin{pmatrix} * & * & 0 \\ * & * & * \\ 0 & * & * \end{pmatrix}$ \\ \noalign{\vskip 1mm} \hline \noalign{\vskip 1mm}
         (6) & $Z_{11}$ & $\{[g^0], [g^4], [g^2]\}$ & $\{[g^3], [g^1], [g^5] \}$ & $[g^1]$ & $\begin{pmatrix} 0 & * & 0 \\ * & 0 & * \\ * & * & 0  \end{pmatrix}$ & $\begin{pmatrix} * & 0 & * \\ 0 & * & * \\ * & * & * \end{pmatrix}$ \\ \noalign{\vskip 1mm} \hline \noalign{\vskip 1mm}
         (7) & $Z_{11}$ & $\{[g^0], [g^1], [g^5]\}$ & $\{[g^3], [g^4], [g^2] \}$ & $[g^3]$ & $\begin{pmatrix} * & 0 & 0 \\ 0 & * & * \\ * & 0 & *  \end{pmatrix}$ & $\begin{pmatrix} * & 0 & * \\ 0 & * & * \\ * & * & * \end{pmatrix}$ \\ \noalign{\vskip 1mm} \hline \noalign{\vskip 1mm}
         (8) & $Z_{11}$ & $\{[g^0], [g^3], [g^5]\}$ & $\{[g^1], [g^4], [g^2] \}$ & $[g^4]$ & $\begin{pmatrix} 0 & * & 0 \\ * & * & 0 \\ * & 0 & *  \end{pmatrix}$ & $\begin{pmatrix} * & * & 0 \\ * & * & * \\ 0 & * & * \end{pmatrix}$ \\ \noalign{\vskip 1mm} \hline
    \end{tabular}
    \caption{Summary of all the viable models with assignment of classes and structures of $y^\eta$ and $m_\nu$. Note that we have other models for each model class by applying permutation within $\{[g^k], [g^{k'}], [g^{k''}]\}$ for $L(\bar{\ell})$. }
    \label{tab:viable-models}
\end{table*}

\section{Analytic formulas for one-zero textures in the model} \label{sec:appendix3}
In this appendix, we summarize analytic relations for one-zero textures of neutrino mass matrix in Eq.~\eqref{eq:oneZero}.
In our notation, the neutrino mass matrix $m_\nu$ can be written by mixing angles, masses, and phases using the relation $m_\nu = U_\nu^* m^d_\nu U_\nu^\dagger$ where $m_\nu^d = {\rm diag}(m_1, m_2 e^{\alpha_{21}/2}, m_3 e^{\alpha_{31}/2})$ and $U_\nu$ is the PMNS matrix. We then obtain the relations associated with one-zero textures as follows; more details can be referred to ref.~\cite{Lashin:2011dn}. 

\noindent
(I) $(m_\nu)_{12} =0$ \\
In this case, we obtain the relation 
\begin{align}
&  -c_{12} c_{13} (s_{12} c_{23} + s_{13} s_{23} c_{12} e^{-i \delta_{\rm CP} }) m_1 \nonumber \\
& + s_{12} c_{13} e^{i \alpha_{21}} (c_{12} c_{23} - s_{12} s_{13} s_{23} e^{-i \delta_{\rm CP}}) m_2 \nonumber \\
& + s_{13} s_{23} c_{13} e^{i(\alpha_{31} + \delta_{\rm CP})} m_3 = 0. \label{eq:oneZero-const1}
\end{align}
The relation is simplified when the neutrino mass matrix is rank 2 as realized in model classes (1) and (2).
For NH case ($m_1 =0$), we have 
\begin{equation}
\frac{m_2}{m_3} =  \frac{-s_{13} s_{23}  e^{i(\alpha_{31} - \alpha_{21} + \delta_{\rm CP})} }{s_{12} (c_{12} c_{23} - s_{12} s_{13} s_{23} e^{-i \delta_{\rm CP}})}.
\label{eq:oneZero-const2}
\end{equation}
For IH case ($m_3 =0$), we have
\begin{equation}
\frac{m_1}{m_2} = \frac{s_{12}  e^{i \alpha_{21}} (c_{12} c_{23} - s_{12} s_{13} s_{23} e^{-i \delta_{\rm CP}})}{c_{12}  (s_{12} c_{23} + s_{13} s_{23} c_{12} e^{-i \delta_{\rm CP} })}.
\label{eq:oneZero-const3}
\end{equation}

\noindent
(II) $(m_\nu)_{13} =0$ \\
In this case, we obtain the relation
\begin{align}
&  c_{12} c_{13} (s_{12} s_{23} - c_{12} c_{23} s_{13} e^{-i \delta_{\rm CP} }) m_1 \nonumber \\
& - s_{12} c_{13} e^{i \alpha_{21}} (s_{23} c_{12} + s_{12} s_{13} c_{23} e^{-i \delta_{\rm CP}}) m_2 \nonumber \\
& + s_{13} c_{13} c_{23} e^{i(\alpha_{31} + \delta_{\rm CP})} m_3 = 0.
\label{eq:oneZero-const4}
\end{align}
We also write relations for rank 2 neutrino mass matrix case.
For NH ($m_1 =0$), we have 
\begin{equation}
\frac{m_2}{m_3} = \frac{s_{13} c_{23} e^{i(\alpha_{31} + \delta_{\rm CP})} }{s_{12} e^{i \alpha_{21}} (s_{23} c_{12} + s_{12} s_{13} c_{23} e^{-i \delta_{\rm CP}})}.
\label{eq:oneZero-const5}
\end{equation}
For IH ($m_3 = 0$), we have 
\begin{equation}
\frac{m_1}{m_2} = \frac{s_{12}  e^{i \alpha_{21}} (s_{23} c_{12} + s_{12} s_{13} c_{23} e^{-i \delta_{\rm CP}})}{c_{12}  (s_{12} s_{23} - c_{12} c_{23} s_{13} e^{-i \delta_{\rm CP} })}.
\label{eq:oneZero-const6}
\end{equation}

\noindent
(III) $(m_\nu)_{23} =0$ \\
In this case, we obtain the relation
\begin{align}
& - (s_{12} s_{23}-c_{12} c_{23} e^{-i \delta_{\rm CP}})(s_{12}c_{23} + s_{13} s_{23} c_{12} e^{-i \delta_{\rm CP}}) m_1 \nonumber \\
& - e^{i \alpha_{21}} (s_{23} c_{12} + s_{12} s_{13} c_{23} e^{-i \delta_{\rm CP}}) \nonumber \\
&\times (c_{12} c_{23} - s_{12}s_{23} c_{13} e^{-i \delta_{\rm CP}} ) m_2 + c_{23} c_{13}^2 s_{23} e^{i \alpha_{13}} m_3 =0.
\label{eq:oneZero-const7}
\end{align}
We also write relations for rank 2 neutrino mass matrix case.
For NH ($m_1 =0$), we have 
\begin{equation}
\frac{m_2}{m_3} = \frac{s_{23} c_{13}^2 s_{23} e^{i( \alpha_{13}-\alpha_{12})}}{ (s_{23} c_{12} + s_{12} s_{13} c_{23} e^{-i \delta_{\rm CP}})(c_{12} c_{23} - s_{12}s_{23} c_{13} e^{-i \delta_{\rm CP}} )}.
\label{eq:oneZero-const8}
\end{equation}
For IH ($m_3 =0$), we have 
\begin{align}
& \frac{m_1}{m_2} \nonumber \\
&= \frac{e^{i \alpha_{21}} (s_{23} c_{12} + s_{12} s_{13} c_{23} e^{-i \delta_{\rm CP}})(c_{12} c_{23} - s_{12}s_{23} c_{13} e^{-i \delta_{\rm CP}} )}{(s_{12} s_{23}-c_{12} c_{23} e^{-i \delta_{\rm CP}})(s_{12}c_{23} + s_{13} s_{23} c_{12} e^{-i \delta_{\rm CP}}) }.
\label{eq:oneZero-const9}
\end{align}

For instance, take BM model 1, where $(m_{\nu})_{13}=0$, leading us to eq.~\eqref{eq:oneZero-const4}, then if we focus on IH ($m_3=0$ since the lightest neutrino mass eigenvalue is zero for BM1), therefore we end up with eq.~\eqref{eq:oneZero-const6}. Finally, plugging in experimental constraints for mixing angles and neutrino mass splittings gives us an exact value solution for the Dirac and Majorana phases, which matches the first (left most) plot in Fig.~\ref{fig:angle1}. Other correlations for Dirac and Majorana phases can be obtained in a similar manner.

%

%
\bibliography{references}
\end{document}